\documentclass{iau}
\usepackage{graphicx}

\title[JD 11.~~Chemistry of FUors] %% give here short title %%
{Chemical modeling of FU~Ori protoplanetary disks}

\author[Tamara Molyarova et al.]   %% give here short author list %%
{Tamara Molyarova$^1$, Vitaly Akimkin$^1$, Dmitry~Semenov$^{2}$, P\'{e}ter~\'{A}brah\'{a}m$^3$, Thomas~Henning$^2$, \'{A}gnes~K\'{o}sp\'{a}l$^{2,3}$, Eduard~Vorobyov$^{4}$, Dmitri~Wiebe$^1$}

\affiliation{$^1$Institute of Astronomy, Russian Academy of Sciences, 119017, 48 Pyatnitskaya st., Moscow, Russia email: {\tt molyarova@inasan.ru} \\
$^2$Max Planck Institute for Astronomy, K{\"o}nigstuhl 17, 69117 Heidelberg, Germany\\
$^2$Konkoly Observatory, Research Centre for Astronomy and Earth Sciences, Hungarian Academy of Sciences, Konkoly-Thege Mikl{\'o}s {\'u}t 15-17, 1121 Budapest, Hungary\\
$^4$Research Institute of Physics, Southern Federal University, Stachki 194, Rostov-on-Don, 344090, Russia}

\pubyear{2019}
\volume{345}  %% insert here IAU Symposium No.
\jname{Origins: from the Protosun to the First Steps of Life}
\editors{Bruce G. Elmegreen, L. Viktor T\'oth, Manuel G\"udel, eds.}
\begin{document}

\maketitle

\begin{abstract}
Luminosity outbursts of the FU~Ori type stars, which have a magnitude of $\sim100$\,$L_{\odot}$ and last for decades, may affect chemical composition of the surrounding protoplanetary disk. Using astrochemical modeling we analyze the changes induced by the outburst and search for species sensitive to the luminosity rise. Some changes in the disk molecular composition appear not only during the outburst itself but can also retain for decades after the end of the outburst. We analyze main chemical processes responsible for these effects and assess timescales at which chemically inert species return to the pre-outburst abundances.
\end{abstract}

Young stars of the FU~Ori type are objects experiencing sudden luminosity outburst, associated with increased accretion rate (\cite[Dunham \& Vorobyov 2012]{DV2012}). These outbursts are thought to be episodic and could happen to many young protostars (\cite[Audard et al. 2014]{Audard14}). Anomalous abundances of specific molecules in disks around quiescent stars may indicate past luminosity outbursts \cite[Rab et al.(2017)]{Rab17}. Our goal is to point out the chemical species useful for this purpose.

\begin{figure}[h]
\begin{center}
 \includegraphics[width=1.0\columnwidth]{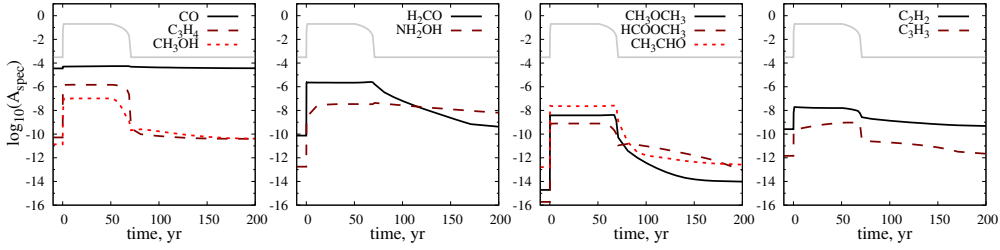}
 \caption{Disk total abundances (with respect to H$_2$) of selected species changing with time. The solid gray line denotes the luminosity profile (in $1000$\,$L_{\odot}$).}
   \label{fig1}
\end{center}
\end{figure}

We use ANDES astrochemical model by \cite[Akimkin et al. (2013)]{Akimkin13}, adjusted to the case of variable accretion rate. We consider disk with the mass $M_{\rm disk} = 0.01$\,$M_{\odot}$ and characteristic radius $R_{\rm c} = 100$\,au around a young sun-like star ($1$\,$M_{\odot}$, $0.9$\,$L_{\odot}$). Dust size distribution is described by a power-law with the index of $-3.5$, $a_{\rm min}=5\cdot10^{-7}$\,cm and $a_{\rm max}=2.5\cdot10^{-3}$\,cm. The outburst is simulated as an increased accretion luminosity lasting for $\approx50$\,yr, with the profile shown in Fig.~\ref{fig1}. The detailed description of this model as well as chemistry of different disk models can be found in \cite[Molyarova et al. (2018)]{Molyarova18}.

The patterns of chemical behavior during the outburst are shown in Figure~\ref{fig1}. Most of gas-phase species grow in abundance during the outburst due to the thermal evaporation of their ices (the first three panels). For CO the increase is not very significant due to the chemical depletion of its ice (\cite[Molyarova et al. 2017]{Molyarova17}, \cite[Bosman et al. 2018]{Bosman18}). Abundant volatiles, such as CH$_3$OH and C$_3$H$_4$, follow closely the luminosity profile and immediately return to their quiescent abundances after the end of the outburst. The species shown in the last three panels of Fig.~\ref{fig1} are found to stay overabundant after the outburst for the longest timescales, from decades to thousand years. 

\begin{figure}[h]
\begin{center}
 \includegraphics[width=1.0\columnwidth]{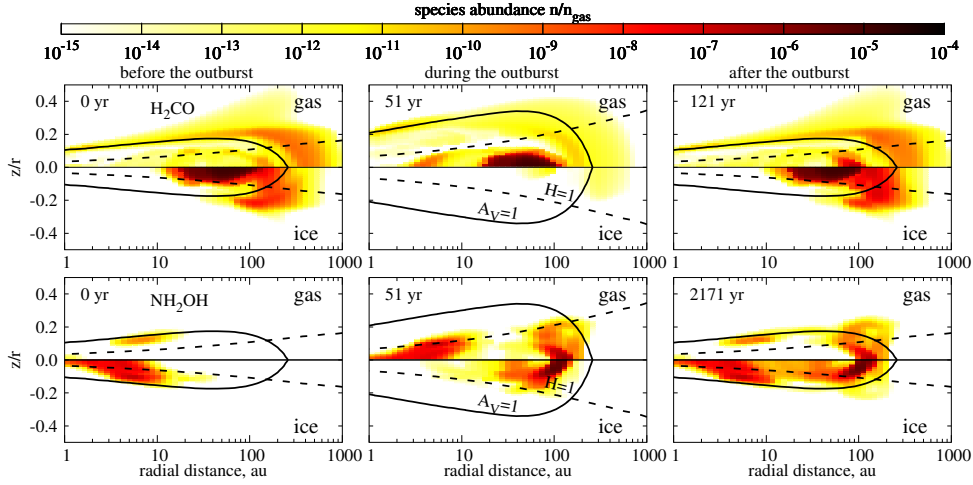}
 \caption{2D distribution of H$_2$CO and NH$_2$OH in gas (upper half-plots) and in ice (lower half-plots) at specified time moments after the beginning of the outburst. Solid and dash black lines show the surface of $A_{\rm v}=1$ and scale height equal to unity, correspondingly.}
   \label{fig2}
\end{center}
\end{figure}

Different mechanisms are responsible for time-dependent chemistry of the selected species. While H$_2$CO, NH$_2$OH, CH$_3$CHO, HCOOCH$_3$, and CH$_3$OCH$_3$ are evaporated from ice, C$_3$H$_3$ and C$_2$H$_2$ are formed directly in the gas. Fig.~\ref{fig2} shows spatial distribution of H$_2$CO and NH$_2$OH, which are the most promising post-outburst tracers. Their abundances are determined mostly by the evaporation, although NH$_2$OH is also actively formed on ice during the outburst. The times of returning to the quiescent values are defined by freeze-out timescales, which are long in the outer disk regions.

The abundances of H$_2$CO and NH$_2$OH jump for several orders of magnitude during the outburst and stay elevated for $10^2-10^3$\,yr after it. This makes them the most suitable candidates for recent outburst tracers.

\textbf{Acknowledgment.} The study is supported by RFBR grant 17-02-00644.


\begin{thebibliography}{}

\bibitem[Akimkin et al.(2013)]{Akimkin13}
{Akimkin, V., Zhukovska, S., Wiebe, D., et al.} 2013, 
\textit{ApJ}, 766, 8


\bibitem[Audard et al.(2014)]{Audard14}
{Audard, M., \'{A}brah\'{a}m, P., Dunham, M. M., et al.} 2014,
\textit{Protostars and Planets VI}, 387

\bibitem[Bosman et al.(2018)]{Bosman18}
{Bosman, A.~D., Walsh, C., and van Dishoeck, E.~F.} 2018,
\textit{A\&A}, 618, A182

\bibitem[Dunham \& Vorobyov (2012)]{DV2012}
{Dunham, M. M., \& Vorobyov, E. I.} 2012,
\textit{ApJ}, 747, 52

\bibitem[Molyarova et al.(2017)]{Molyarova17}
{Molyarova, T., Akimkin, V., Semenov, D., et al.} 2017,
\textit{ApJ}, 849, 130

\bibitem[Molyarova et al.(2018)]{Molyarova18}
{Molyarova, T., Akimkin, V., Semenov, D., et al.} 2018,
\textit{ApJ}, 866, 46

\bibitem[Rab et al.(2017)]{Rab17}
{Rab, C., Elbakyan, V., Vorobyov, E., et al.} 2017,
\textit{A\&A}, 604, A15


% more references

\end{thebibliography}
\end{document}